\documentclass[12pt, twoside]{article}
\usepackage{a4wide,amssymb,cite}
\usepackage{epsfig}
\usepackage[usenames,dvipsnames]{color}
\usepackage{slashed}

\newcommand{\be}{\begin{equation}}
\newcommand{\ee}{\end{equation}}
\newcommand{\bea}{\begin{eqnarray}}
\newcommand{\eea}{\end{eqnarray}}
\def\eq#1{eq.~(\ref{#1})}
\def\fig#1{fig.~\ref{#1}}
\def\bma#1{\mbox{\boldmath{$#1$}}}
\def\simlt{\stackrel{<}{{}_\sim}}
\def\simgt{\stackrel{>}{{}_\sim}}

\begin{document}

\color{black}
\begin{flushright}
CERN-TH/2012--059\\
\end{flushright}

\vspace{1cm}
\begin{center}
{\Huge\bf\color{black} Stabilization of the Electroweak\\[3mm] Vacuum by a Scalar Threshold Effect}\\
\bigskip\color{black}\vspace{0.6cm}{
{\large\bf  Joan Elias-Mir\'o$^a$, Jos\'e R. Espinosa$^{a,b}$, Gian F. Giudice$^{c}$,\\ [2mm]  Hyun Min Lee$^{c}$, Alessandro Strumia$^{d,e}$}
\vspace{0.5cm}
} \\[7mm]
{\em $(a)$ {IFAE and Physics Dept., Univ. Aut\`onoma de Barcelona, Barcelona, Spain}}\\
{\em $(b)$ {ICREA, Instituci\`o Catalana de Recerca i Estudis Avan\c{c}ats, Barcelona, Spain}}\\
{\em $(c)$ {CERN, Theory Division, CH--1211 Geneva 23,  Switzerland}}\\
{\it $(d)$ Dipartimento di Fisica dell'Universit{\`a} di Pisa and INFN, Italy}\\
{\it  $(e)$ National Institute of Chemical Physics and Biophysics, Estonia}\\
\end{center}
\bigskip
\centerline{\large\bf Abstract}
\begin{quote}\large
We show how a heavy scalar singlet with a large vacuum expectation value can
evade the potential instability of the Standard Model electroweak vacuum. 
The quartic interaction between the heavy scalar singlet and the Higgs doublet leads to a positive tree-level threshold correction for the Higgs quartic coupling, which is very effective in stabilizing the potential. 
We provide examples, such as the see-saw,  invisible axion and unitarized Higgs inflation, where the proposed mechanism is automatically implemented in well-defined ranges of Higgs masses.

\end{quote}

\thispagestyle{empty}

\normalsize

\newpage

\setcounter{page}{1}

\section{Introduction}
Collider searches have restricted the allowed mass window for the Standard Model Higgs boson to the narrow range between 115.5 and 127 GeV at 95\% CL~\cite{ATLASCMS}. From the theoretical point of view, this is an interesting range of masses because in this interval the Higgs potential can develop an instability at large field values \cite{EEGIRS,Lind}. This is of course not necessarily a problem, first because the Standard Model is likely to be embedded in a more fundamental theory at high energies which may change the shape of the Higgs potential and, second, because the actual fate of the electroweak vacuum depends on the cosmological history. It should be remarked, however, that an instability caused by a negative quartic coupling is not cured at high temperatures (as opposed to an instability caused by a negative quadratic term) and typically persists in the early universe. Therefore, to avoid potential cosmological constraints it may be preferable to cure any Higgs instability at large field values.

\smallskip

There are, of course, infinitely many ways to modify the Higgs potential and raise the instability scale $\Lambda_I$. 
In this paper we point out a very simple and economical mechanism to avoid an instability of the electroweak vacuum. It requires the existence of a new heavy scalar singlet that acquires a large vacuum expectation value (vev) and has a quartic interaction with the ordinary Higgs doublet. The crucial point is that the matching condition of the Higgs quartic coupling, at the scale where the singlet is integrated out, corresponds to a positive shift, as we evolve from low to high energies. Although the stability condition is also modified by the presence of the singlet, a careful analysis shows that, under the conditions specified in section~2, the threshold correction  helps to stabilize the potential. The effect occurs at tree level and thus can be sizable and, in general, dominant over loop contributions. Moreover, the effect does not decouple, in the sense that the size of the shift does not depend on the singlet mass, which could take any value lower than the instability scale $\Lambda_I$. 

The dynamics of the mechanism is explained in section~\ref{sec2}.
In section~\ref{sec3} we put the proposed mechanism into context, considering setups in which the existence of a heavy scalar singlet has independent motivations. We present three examples: see-saw origin of neutrino masses, invisible axion, and Higgs inflation. For each example we give the range of Higgs masses where the corresponding scalar field could cure the instability. 

\section{Stabilizing the Higgs with a Scalar Singlet}
\label{sec2}

To explore the impact 
of an additional singlet scalar on the stability of the Higgs potential, we
consider a tree-level scalar potential of the form  
\be
V_0=\lambda_H \left( H^\dagger H -v^2/2\right)^2 +\lambda_S  \left( S^\dagger S -w^2/2\right)^2
+2\lambda_{HS}  \left( H^\dagger H -v^2/2\right)  \left( S^\dagger S -w^2/2\right) \ . 
\label{pot}
\ee
Here $H$ is the Higgs doublet, $S$ is a complex scalar field, and $V_0$ is the most general renormalizable potential that respects a global abelian symmetry under which only $S$ is charged. Although we will consider here a single complex scalar, most of our conclusions remain valid also in the case of multi-Higgs doublets or real singlet fields (with a $Z_2$ parity replacing the abelian symmetry). 

\smallskip

For $\lambda_H, \lambda_S>0$ and $\lambda_{HS}^2<\lambda_H \lambda_S$, the minimum of $V_0$ is at
\be
\langle H^\dagger H \rangle =v^2/2\ ,~~~~\langle S^\dagger S \rangle =w^2/2\ .
\ee
A nonzero vev of $S$, which is crucial for our mechanism to work, spontaneously breaks the global symmetry (or the $Z_2$ parity, for a real singlet) giving rise to a potentially dangerous Goldstone boson (or domain walls). Gauging the symmetry of $S$ or explicitly breaking it
by (possibly small) terms in $V_0$ can be used to evade these problems, but does not conceptually modify our results. For simplicity, we restrict our considerations to the potential in eq.~(\ref{pot}), but generalizations are straightforward.

The presence of the new scalar field $S$ modifies the analysis of the stability conditions of the Higgs potential. 
One  effect, already considered in previous literature is
the contribution of the singlet to the renormalization group evolution of the
Higgs quartic coupling (for recent analyses, see \cite{RGEsing} and references therein). The relevant renormalization group equations (RGEs) above the scale $M_S=\sqrt{2\lambda_S}w$ are, at 
one-loop:
\begin{eqnarray}
(4\pi)^2\frac{d\lambda _{{H}}}{d\ln\mu} &=& \left(12 y_t^2-3 {g'}^2-9 g^2\right) \lambda _{{H}}-6 y_t^4+\frac{3}{8}\left[2g^4+ ({g'}^2+g^2)^2\right]+24 \lambda _{{H}}^2+4 \lambda _{{HS}}^2 \ , \nonumber \\
(4\pi)^2\frac{d\lambda _{{HS}}}{d\ln\mu} &=& \frac{1}{2}\left(12 y_t^2-3 {g'}^2-9 g^2\right) \lambda _{{HS}}+
4 \lambda _{{HS}}\left(3 \lambda _{{H}}+2 \lambda _{{S}}\right)+8 \lambda _{{HS}}^2\ ,   \label{RG} \\
(4\pi)^2\frac{d\lambda _{{S}}}{d\ln\mu} &=& 8 \lambda _{{HS}}^2+20 \lambda _{{S}}^2\ .  \nonumber
\end{eqnarray}
If the singlet mass $M_S$ is below the SM instability
scale\footnote{Stabilizing the potential with degrees of freedom heavier than $\Lambda_I$ requires sizable couplings, see \cite{strong}.} $\Lambda_I$ and $(\lambda_{HS}/4\pi)^2\ln(\Lambda_I/M_S)$ is large enough, the positive contribution to the RGE equation for
$\lambda_H$ can prevent it from becoming negative.

\begin{figure}
$$\includegraphics[width=0.50\textwidth]{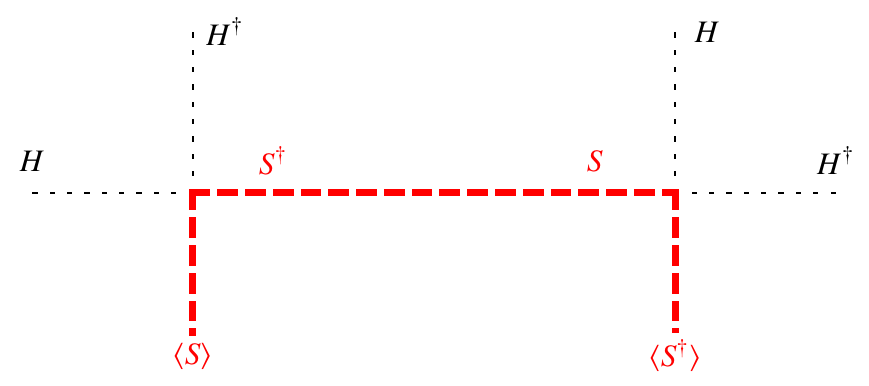}$$ 
\caption{\em Feynman diagram for the tree-level threshold correction to the Higgs quartic coupling.
\label{Feyn}}
\end{figure}

\subsection{Threshold Effect}
Besides the loop contribution discussed above, there is a related tree-level effect through which the new singlet 
can affect the stability bound. Let us consider the limit in which $M_S$ is much larger than the Higgs mass ($w^2\gg v^2$). At the scale $M_S$ we can integrate out the field $S$ using its equation of motion (neglecting derivatives):
\be
S^\dagger S = {w^2\over 2} -\frac{\lambda_{HS}}{\lambda_S} \left( H^\dagger H -\frac{v^2}{2}\right).
\label{eqmot}
\ee
Replacing eq.~(\ref{eqmot}) in $V_0$, we obtain the effective potential below the scale $M_S$:
\be
V_{\rm eff} =\lambda \left( H^\dagger H -{v^2\over 2} \right)^2,~~~~~\lambda =\lambda_H-\frac{\lambda_{HS}^2}{\lambda_S}\  .
\label{match}
\ee
This shows that the matching condition at the scale $Q=M_S$ of the Higgs quartic coupling gives a tree-level shift, $\delta\lambda\equiv \lambda_{HS}^2/\lambda_S$, as we go from $\lambda_H$ just above $M_S$ to $\lambda$ just below $M_S$. 
Figure~\ref{Feyn} shows the Feynman diagram that gives rise to such tree-level shift.

To better understand the origin of the shift in the matching condition, let us consider the mass matrix of the fields $h$ and $s$, corresponding to the real parts of the doublet $H$ (in unitary gauge) and the singlet $S$, such that $H^\dagger H=h^2/2$ and $S^\dagger S=s^2/2$. At the minimum, the mass matrix is
\be
{\cal M}^2=2\left(\begin{array}{cc}
\lambda_H v^2 & \lambda_{HS} v w\\
\lambda_{HS} v w & \lambda_S w^2
\end{array}\right)\ .
\ee
In the limit $\lambda_S w^2\gg \lambda_H v^2$, the heaviest eigenstate, which is nearly singlet, can be integrated out, leaving behind a ``see-saw"-like correction to the lightest eigenvalue
\be
\label{NewMh}
m_h^2 = 2 v^2\left[ \lambda_H - \frac{\lambda_{HS}^2}{\lambda_S}+{\cal O}\left(\frac{v^2}{w^2}\right) \right] \ ,
\ee
while $M_S^2=2\lambda_S w^2+2(\lambda_{HS}^2/\lambda_S)v^2+{\cal O}(v^4/w^2)$.
The light state is almost purely $h$, as the singlet admixture is suppressed by a small mixing
angle of order $v/w$. However, the Higgs mass correction due to the heavy state persists even in the decoupling limit ($w \to \infty$). The negative sign in the shift of the Higgs mass in eq.~(\ref{NewMh}) can be readily understood as coming from the repulsion of mass eigenvalues after turning on the mixing equal to $2 \lambda_{HS}vw$.

\begin{figure}[t]
$$\includegraphics[width=0.45\textwidth]{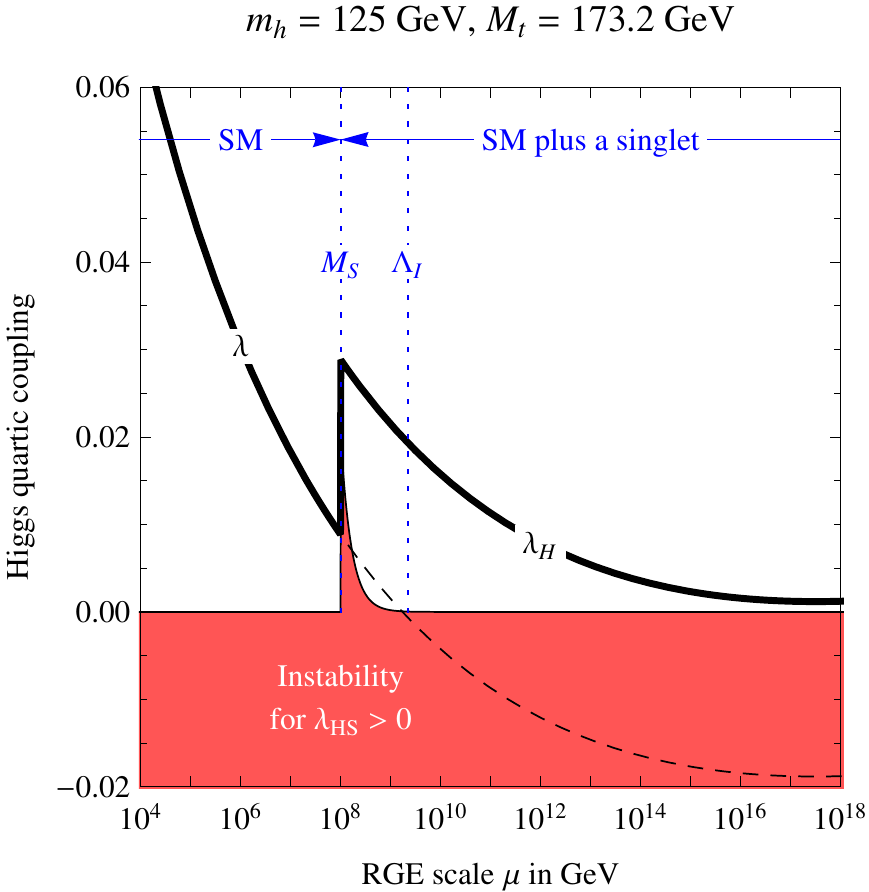}\qquad\includegraphics[width=0.45\textwidth]{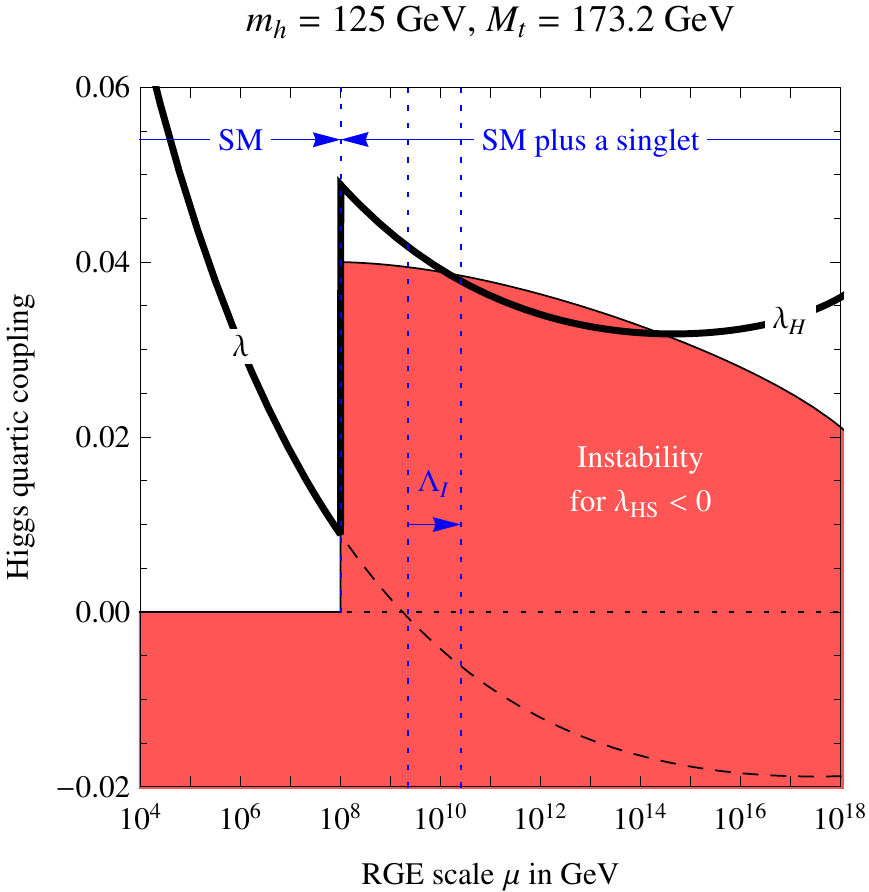}$$ 
\caption{\em Running of the Higgs quartic coupling in the SM and in the model with a scalar singlet,
here assumed to have the mass $M_S =10^8\,{\rm GeV}$.
Left: if $\lambda_{HS}>0$, thanks to the tree level shift at the singlet mass, the coupling never enters into the instability region, even assuming that singlet contributions to the RG equations are negligible.
Right: if $\lambda_{HS}<0$ the instability can be shifted away or avoided only by
singlet contributions to the RG equations.
\label{run}}
\end{figure}

\bigskip

Naively, as the tree-level shift $\delta\lambda$ corresponds to a larger Higgs quartic coupling above $M_S$, the chances of keeping it positive seem improved. However, the tree-level conditions for stability change from $\lambda >0$ in the effective theory below $M_S$ to $\lambda_H >\delta \lambda$ in the full theory above $M_S$. Thus, it appears that the threshold correction $\delta \lambda$ does not help stability at all. To understand what happens, one has to reexamine the stability conditions more carefully.
First of all, remember that the tree-level potential $V_0$ in eq.~(\ref{pot}) is a good approximation to the full potential if we evaluate couplings and masses (collectively denoted by $\lambda_i$ below)  at a renormalization scale of the order of the field values of interest. Once we express the scalar potential as $V_0[\lambda_i(\mu=\varphi),\varphi]$, potentially large logarithms of the form $\ln m_i(\varphi)/\mu$ (where $m_i(\varphi)\sim\varphi$ is a typical field-dependent mass) are kept small. Roughly speaking, this means that  $V_0$  with a fixed $\mu_c$ will be reliable as long as one examines $\varphi\sim\mu_c$ and restricts field excursions to $|\varphi - \mu_c|< \mu_c e^{8\pi^2\lambda_0/\lambda_1^2}$ (where $\lambda_0$ denotes a coupling in the tree-level potential
and $\lambda_1$ a coupling affecting the radiative corrections, e.g. the top Yukawa coupling squared). By adjusting $\mu\sim\varphi$ one can evaluate reliably the potential at all field values, but the previous estimate tells us when we can use $V_0[\lambda_i(\mu_c),\varphi]$, which has a simpler field dependence. 

With the parametrization chosen in eq.~(\ref{pot}), the EW vacuum corresponds to $V_0 =0$. Thus the stability condition is $V_0>0$ anywhere in field space, away from the EW vacuum. 
The first stability requirement that we should impose is
\be\label{stab1}
\lambda_H(\mu)>0\ ,\quad\lambda_S(\mu)>0\ ,
\ee
at any renormalization scale $\mu$, or else the potential develops unwanted minima lower than the EW vacuum or is unbounded from below at large field values. 

Next, in order to discuss the conditions on the coupling constant $\lambda_{HS}$, it is convenient to separate the cases in which $\lambda_{HS}$ is either positive or negative. This separation is meaningful because $\lambda_{HS}$ renormalizes multiplicatively (as it is the only coupling that connects $H$ and $S$), see eq.~(\ref{RG}), and therefore the RG flow cannot flip its sign. 

\medskip

\subsection{Case ${\bma{\lambda_{HS}>0}}$}
In this case, $V_0$ can become negative only when $|S|<w/\sqrt{2}$ (neglecting corrections proportional to $v$). In this situation, the most dangerous field configuration is 
well approximated by setting $S=0$ in \eq{pot}, such that
\be
V_0(H,0)\approx \lambda_H |H|^4 -\frac{\lambda_{HS}}{2\lambda_S}M^2_S|H|^2 +\frac{M^4_S}{16\lambda_S}\ .
\ee  
The extra stability condition ($V_0>0$) is then
\be
\lambda_{HS}^2 (\mu ) <  \lambda_H(\mu) \lambda_S (\mu)\ .
\label{condp}\label{stabplus}
\ee
Note that  this can be rewritten as $\lambda_H>\delta\lambda=\lambda_{HS}^2/\lambda_S$
and ensures that the light scalar state does not become tachyonic, see eq.~(\ref{NewMh}). 
If this condition were violated at some scale $\mu_*$, it would lead to an instability for field configurations with
\be
|S|<\frac{M_S}{2\sqrt{\lambda_S}}, ~~~\mu_- <|H| < \mu_+, ~~~\mu_\pm^2= \frac{M_S^2\lambda_{HS}}{4\lambda_H\lambda_S} \left. \left(1\pm\sqrt{1-\frac{\lambda_H\lambda_S}{\lambda^2_{HS}}} \right) \right|_{\mu_*}\ ,
\ee
which could be trusted provided $\mu_-<\mu_*<\mu_+$. Note that, if $\mu_*\gg \mu_\pm$, this would not mean that there is an instability to worry about, as it would be located outside the range of validity of the
tree-level approximation $V_0(\lambda_i(\mu_*),\varphi)$.
Thus, as long as condition (\ref{condp}) is satisfied for renormalization scales within a relatively narrow range of energies around $M_S$ (which fixes the mass scale of $\mu_\pm$), there is no instability 
even if this condition were eventually violated at higher scales. 
Only if parameters happen to lie near a critical point in which at least one of conditions (\ref{stab1}) or (\ref{condp}) is barely satisfied, radiative corrections can become important and invalidate the stability analysis performed with the tree-level potential. In this case one should resort to the one-loop approximation of the potential; otherwise, our analysis is reliable.

\medskip

We can now better appreciate how the threshold contribution in \eq{match} can cure the instability of the SM Higgs potential (provided that $M_S<\Lambda_I$). The correction $\delta \lambda$ has the correct sign to shift the Higgs quartic coupling upwards ($\lambda_H=\lambda +\delta \lambda$), although the stability condition is also shifted upwards by the same amount, becoming $\lambda_H > \delta \lambda$. However, for positive $\lambda_{HS}$, the condition 
$\lambda_H > \delta \lambda$ has to be satisfied only at scales of order $M_S$, while for larger scales it rapidly reduces to the conventional constraint $\lambda_H>0$. Moreover, one-loop RG effects (although typically less important than the tree-level matching condition) also help to maintain stability. First, $\lambda_S$ and $\lambda_{HS}$ will stay positive
once they are positive at $M_S$. Second, $\beta_{\lambda_H}\equiv d\lambda_H/d\ln\mu$ receives extra positive contributions proportional to $ \lambda_{HS}^2$ and  to $\lambda_H^2$ (coupling which is  numerically larger after the threshold shift).  These two RG effects can reduce  (or even overcome) the destabilizing effect from top loops.

\begin{figure}[t]
$$\includegraphics[width=0.70\textwidth]{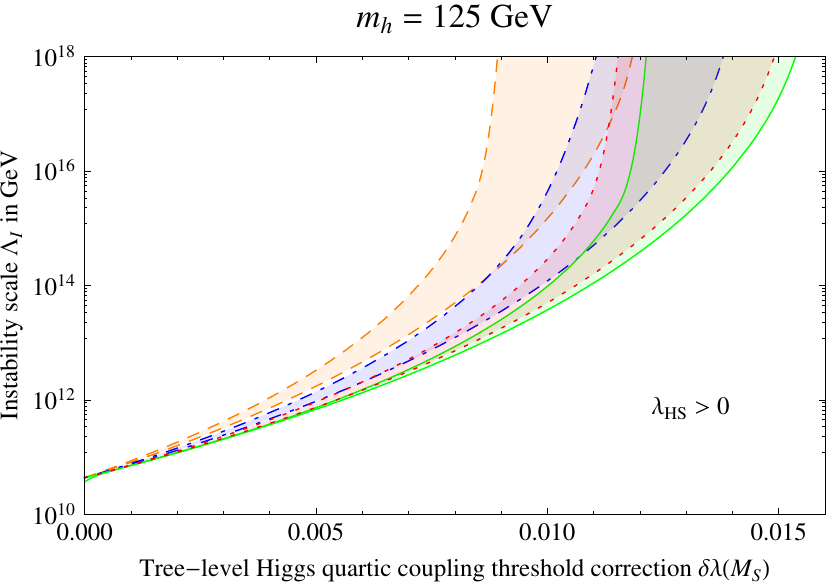}$$ 
\caption{\em For $m_h=125$ {\em GeV} and $\lambda_{HS}>0$, bands of the modified instability scale $\Lambda_I$  versus the threshold correction  $\delta\lambda$ to the Higgs quartic coupling due to a scalar singlet with mass $M_S=10^4, 10^6, 10^8, 10^{10}$ {\rm GeV} (from left to right). For a fixed $M_S$ value the lowest boundary of the band corresponds to small $\lambda_S, \lambda_{HS}$ and the highest boundary to $\lambda_S(M_{\rm Pl})=4\pi$.  
\label{LIshift}}
\end{figure}

To illustrate the situation, we show in \fig{run}, left panel,  how the Higgs quartic coupling runs with the renormalization scale.
We consider $M_S = 10^8$ GeV $\ll \Lambda_I^{\rm SM} = 2\times10^{9}\,{\rm GeV}$.\footnote{Strictly speaking, this is the scale at which $\lambda=0$, and corresponds to the instability scale of the tree-level RG-improved potential. The $\Lambda_I$ that we calculate later on is higher and corresponds to the instability of the one-loop RG-improved potential. For simplicity, in  fig.~\ref{LIshift} we simply plot $\lambda(\mu)$. }
For simplicity we take the couplings of the singlet to be smaller than the SM top and gauge couplings,
in order to better isolate the tree-level threshold effect.
The same panel also shows the full stability condition, computed numerically by demanding that
$V(H,0)>0$: we see that at renormalization scales just above $M_S$ the stability
condition of~\eq{stabplus} matters, but at larger field values it
rapidly becomes irrelevant and only $\lambda_H>0$ remains.

To study the efficiency of the stabilization mechanism, we performed a numerical study using the full one-loop effective potential with SM couplings running at two-loops. We limited the evolution of the unknown singlet couplings to the one-loop level, given that the effect we are considering is at tree level. In order to track accurately the large field behaviour of the one-loop potential one can simply include in the running quartic couplings the finite one-loop contributions not captured by RG evolution and impose the stability conditions on these corrected couplings (the shift in the instability scale can be up to one order of magnitude; see \cite{CEQ,EEGIRS} for further details). The scale at which these one-loop improved running couplings
violate the stability condition corresponds then to the field scale at which the potential falls below the EW vacuum.
The results  are illustrated in fig.~\ref{LIshift} which shows the new instability scale $\Lambda_I$ as a function of the threshold shift $\delta\lambda$ for several singlet scalar masses,
$M_S=10^4, 10^6, 10^8, 10^{10}$ GeV below the SM instability scale, $\Lambda^{\rm SM}_I\simeq 4\times 10^{10}$
GeV (for $m_h=125$ GeV, $M_t=173.2$ GeV and $\alpha_{\rm s}(M_Z)=0.1183$). For each value of $M_S$, there is a band of values for $\Lambda_I$ due to the freedom in choosing $\lambda_S$, once $\lambda_H$ and $\delta\lambda=\lambda_{HS}^2/\lambda_S$ are fixed. The lower boundary of each band corresponds to $\lambda_S\ll 1$ (and consequently also $\lambda_{SH}\ll1$ for fixed and small $\delta \lambda$). This case nearly isolates the impact of the tree-level shift on the instability scale (as the running of $\lambda_H$ above the singlet threshold is SM-like). The upper boundary of each band corresponds to the largest value of $\lambda_S$ that we allow by requiring $\lambda_S(\mu)<4\pi$ up to the Planck scale. Large values of $\lambda_S$ correspond to large $\lambda_{HS}$, making the RG effect on $\lambda_H$ stronger. We conclude that the tree-level shift in $\lambda$ can have an extremely significant impact in raising the instability scale even for very moderate values of the couplings $\lambda_S$ and $\lambda_{HS}$, and it can easily make the EW vacuum absolutely stable.

\subsection{Case ${\bma{\lambda_{HS}<0}}$} 
In this case $V_0$ can become negative only for $|S|>w/\sqrt{2}$. In this condition, we can neglect the mass parameters $v$ and $w$ in eq.~(\ref{pot}) and approximate the potential by keeping only the quartic terms
\be
V_0 \approx  \lambda_H  |H|^4 +\lambda_S |S|^4
+2\lambda_{HS}   |H|^2 |S|^2  \ .
\label{pot4}
\ee
The stability condition ($V_0>0$) is now 
\be
-\lambda_{HS} (\mu ) < \sqrt{ \lambda_H(\mu) \lambda_S (\mu)}\ .
\label{condm}
\ee
If this condition is violated at some scale $\mu_*$ an instability would develop with
\be
|S|>\frac{M_S}{2\sqrt{\lambda_S}}, ~~~ c_-  < \frac{|H|}{|S|} < c_+ ~~~c_\pm^2= \frac{-\lambda_{HS}}{\lambda_H}\left. \left(1\pm\sqrt{1-\frac{\lambda_H\lambda_S}{\lambda^2_{HS}}}\right) \right|_{\mu_*} .
\label{direc}
\ee
As this determines a direction in field space along which the fields $H$ and $S$ slide towards an unbounded instability, condition (\ref{condm}) has to be satisfied at all renormalization scales larger than $M_S$. Thus the stability condition for negative $\lambda_{HS}$ is much more constraining than in the case of positive $\lambda_{HS}$.

\bigskip

In the case $\lambda_{HS}<0$, as the stability condition $\lambda_H >\delta \lambda$ must be satisfied at all scales, the
tree-level threshold effect is not sufficient to
improve the stability.  Then one should resort to RG effects to improve the potential stability,
as illustrated in fig.~\ref{run}, right panel.
By using the RG equations (\ref{RG}), we can derive the evolution of the effective Higgs quartic coupling
combination $\lambda\equiv \lambda_H-\lambda^2_{HS}/\lambda_S$ above $M_S$ as 
\be
\frac{d\lambda}{d\ln\mu} = \beta^{\rm SM}_\lambda+\frac{8}{(4\pi)^2}\left[(\lambda_{HS}-\delta\lambda)^2 +3\lambda\ \delta\lambda\right]\ ,
\ee
where $\beta^{\rm SM}_\lambda$ is the SM beta function for the Higgs quartic coupling and
$\delta\lambda=\lambda^2_{HS}/\lambda_S>0 $. We see that the additional term in the beta function of $\lambda$ is always positive so that RG effects tend to increase the instability scale also in the case
$\lambda_{HS}<0$. 

\begin{figure}
$$\includegraphics[width=0.70\textwidth]{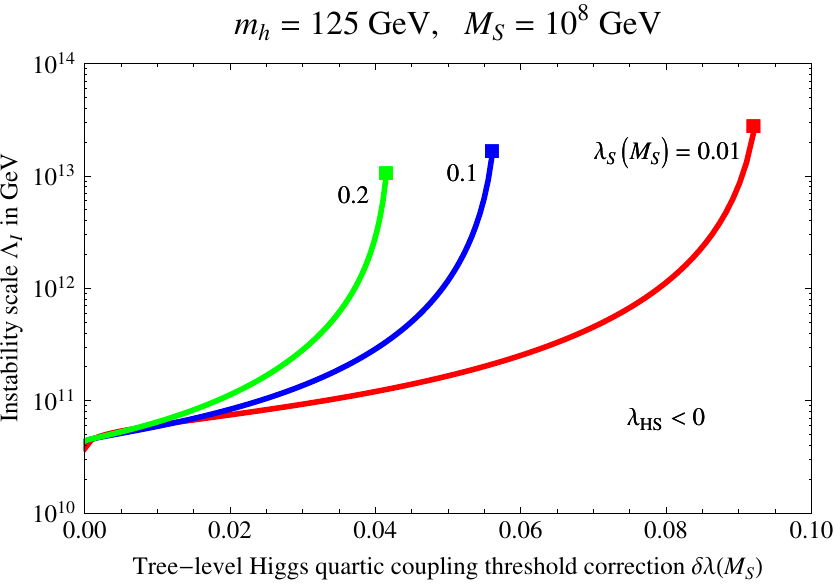}$$ 
\caption{\em For $m_h=125$ {\em GeV} and $\lambda_{HS}<0$, the modified instability scale $\Lambda_I$  versus the threshold correction  $\delta\lambda$ to the Higgs quartic coupling due to a scalar singlet with mass $M_S= 10^8$ {\rm GeV} and  $\lambda_S(M_S)=0.01, 0.1, 0.2$, as indicated.  
\label{LI2shift}}
\end{figure}

The numerical analysis of the $\lambda_{HS}<0$ case confirms this expectation. As an illustration, fig.~\ref{LI2shift} shows the instability scale versus the shift $\delta\lambda$ for the same choice of SM parameters as in fig.~\ref{LIshift} and for the particular case $M_S=10^8$ GeV with three different values of $\lambda_S$ as indicated. The end-point of the curves marks the location beyond which (i.e. for larger $\delta\lambda$) the potential   becomes completely stable. These end-points occur because $\lambda$ first decreases as a function of the renormalization scale but, after reaching a minimum, starts increasing at large scales. In comparison with the case $\lambda_{HS}>0$ (fig.~\ref{LIshift}) we see that larger values of the shift $\delta\lambda$ are now required to have a significant impact on the instability scale.

The stabilization mechanism for $\lambda_{HS}<0$ we have just described is fragile with respect to possible new contributions to the RGEs that can appear if the singlet couples to other sectors of the theory. In contrast, the stabilization mechanism for  $\lambda_{HS}>0$ is more robust, being based on a tree-level shift.
The mechanism is also very effective (because the tree level can be easily large) and economical (because it requires only a heavy scalar singlet). The proposed mechanism can be realized in several situations of physical interest and we now turn to discussing some examples.

\section{Examples}
\label{sec3}

In this section, we present some examples in which a heavy scalar singlet has been advocated as a solution to other SM problems such as the smallness of neutrino masses, strong CP problem and the unitarity problem of Higgs inflation.

\subsection{See-saw}

The see-saw is the conventional mechanism to understand the smallness of neutrino masses. 
It assumes the existence of heavy right-handed neutrino states $N$ with
\be
{\cal L}_N= i\bar{N}\slashed{\partial} N + y_\nu L N H +\frac{M_N}{2} N^2 +{\rm h.c.}
\ee
After EW symmetry breaking, nonzero neutrino masses are generated
\be
m_\nu = \frac{y_\nu^2 v^2}{M_N}\ ,
\ee
which are naturally small provided $M_N\gg v$. 

The impact of the see-saw mechanism on the stability of the Higgs potential has been discussed in the past \cite{stabnu1,stabnu2,EEGIRS}.
The right-handed neutrino Yukawa couplings can play a destabilizing role on $\beta_\lambda$ similar to that of the top Yukawa
coupling. As they scale like $y_\nu^2\sim m_\nu M_N$, they become sizable for large $M_N$ and are dangerous
for stability only if $M_N\simgt 10^{13}$ GeV. For lower $M_N$ the new Yukawas will have a negligible effect on stability.

We do not know what originates the large right-handed Majorana mass, but the simplest idea is to assume that the right-handed neutrinos are coupled to a scalar field carrying two units of lepton number and having a large vev,
\be
\frac{\kappa }{2}\ S \, N^2 +{\rm h.c.}
\ee
The vev of $S$, which sets the scale of the Majorana mass, $M_N=\kappa \langle S\rangle$ does not necessarily lead to a Goldstone boson because in unified models $B-L$ is usually a gauge symmetry. 
In this well-motivated realization of the see-saw the scalar field $S$ could naturally reestablish stability of the electroweak vacuum.  In this setting the role of the singlet scalar is therefore double. Upon taking a large vev and decoupling, it leaves behind two effects: a Weinberg dimension-5 operator that gives neutrinos a nonzero mass and a threshold effect on the Higgs Yukawa coupling that solves the stability problem of the Higgs potential, as long as the mass of $S$ is smaller than $\Lambda_I$.

\begin{figure}[t]
$$\includegraphics[width=0.70\textwidth]{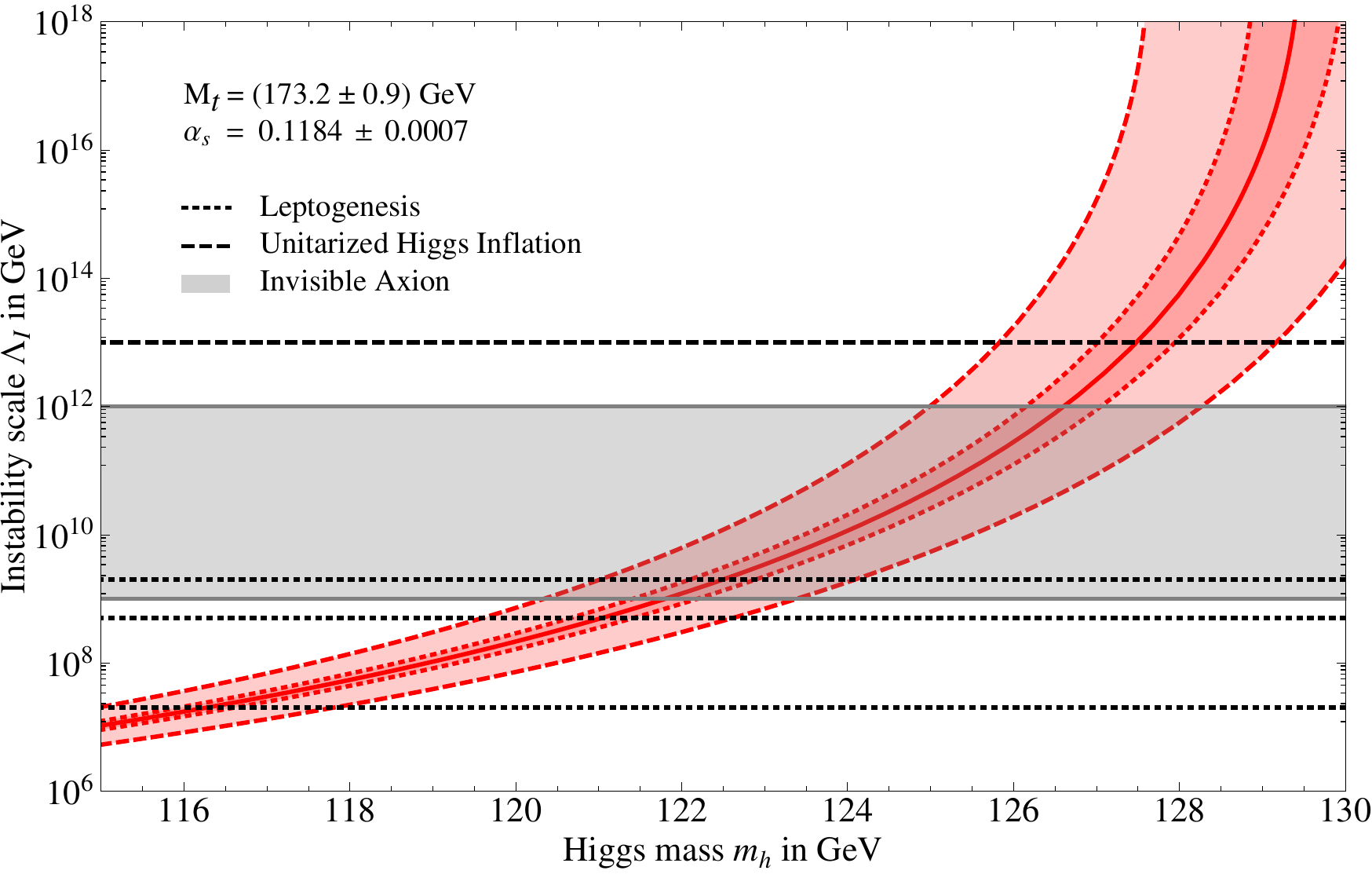}$$ 
\caption{\em The SM instability scale $\Lambda_I$ increasing as a function of the Higgs mass. The central line corresponds to $M_t=173.2$ {\em GeV} and $\alpha_{\rm s}(M_Z)=0.1184$ and the side-bands to 1 sigma deviations as indicated (with the larger deviation for the top mass uncertainty). The horizontal lines and band mark several values or ranges of interest for $\Lambda_I$. The three lowest lines are relevant for the see-saw case and correspond to lower limits on the mass $M_1$ of the lightest right-handed neutrino $N_1$ coming from thermal leptogenesis. The bound depends on the initial density $\rho_{N_1}$:
$M_1> 2\times 10^7$ {\rm GeV} for $\rho_{N_1}\sim 0$; $M_1> 5\times 10^8$ {\rm GeV} for thermal $\rho_{N_1}$ and $M_1>2\times 10^9$ {\rm GeV} for $\rho_{N_1}$ dominating the universe; the shaded band shows the range of singlet masses $10^9-10^{12}$ {\rm GeV} relevant for the axion case; and the upper line is the singlet mass $10^{13}$ {\rm GeV} relevant for the unitarized Higgs inflation case.
\label{SMLIMH}}
\end{figure}

A lower bound on the lightest right-handed neutrino mass $M_1$ is derived by assuming that the cosmic baryon asymmetry is explained by thermal leptogenesis.\footnote{If neutrinos are nearly degenerate in mass, thermal leptogenesis could operate at much smaller values of $M_1$ and the following lower bounds do not apply.} In this case, one obtains the bounds~\cite{neulim}:
\begin{itemize}
\item 
$M_1> 2\times 10^9$ GeV, if the initial right-handed neutrino density vanishes at high temperature.
\item
$M_1> 5\times 10^8$ GeV, if the initial right-handed neutrino density is thermal at high temperature.
\item
$M_1> 2\times 10^7$ GeV, if the initial right-handed neutrino density dominates the universe at high temperature.
\end{itemize}
Assuming that the mass of $S$ is equal or smaller than its vev, we can infer the range of Higgs masses for which the scalar setting the see-saw scale could cure any instability of the potential. From 
fig.~\ref{SMLIMH} we can easily read off such Higgs masses.\footnote{Fig.~\ref{SMLIMH} gives the SM instability scale as a function of $m_h$ as calculated in \cite{EEGIRS}. Besides the 1-sigma error bands shown, associated with the experimental uncertainties in $M_t$ and $\alpha_s$, a (conservative) estimate of the higher order radiative corrections not included in the calculation results in a theoretical uncertainty on $m_h$ of $\pm 3$ GeV \cite{EEGIRS}.} At 90\% CL in $M_t$ and $\alpha_{\rm s}$, we find that the see-saw singlet can potentially eliminate the instability of the EW vacuum, as long as $m_h> 120$~GeV (leptogenesis with vanishing initial right-handed neutrino density), $m_h> 119$~GeV (leptogenesis with thermal initial right-handed neutrino density), or $m_h> 115$~GeV (leptogenesis with dominant initial right-handed neutrino density).
These limits are compatible with the region of Higgs masses suggested by preliminary ATLAS and CMS data, $m_h =124-126$ GeV. The instability scale is raised according to the mechanism discussed in the previous section as long as $M_N\simlt 10^{13}$ GeV,
since the RG effects of $y_\nu$ couplings can be neglected.

We conclude that this simple scenario could comfortably account for the cosmological baryon asymmetry through leptogenesis, for the smallness of neutrino masses and cure the Higgs potential instability. The only drawback of this (beautifully simple but depressing) scenario is that it makes plainly explicit the hierarchy problem: a large singlet vev also 
gives a tree-level contribution to the Higgs mass term in the Lagrangian which requires a large fine-tuning.
(This is in contrast with the scenario without the singlet, in which Higgs mass corrections appear at one-loop and are dangerous only when $M_N>10^7$ GeV \cite{hiernu}).

\subsection{Invisible Axion}
   
The scalar field $S$ can also be identified with the invisible axion.  DFSZ axion models~\cite{DFSZ} use the SM fermion content and a two-Higgs doublet structure $H_u$ and $H_d$
 augmented by a complex scalar $S$, neutral under SM gauge interactions,  
with a coupling $\lambda_{HS} S^2 H_u H_d + \hbox{h.c.}$,
 analogous to the one in eq.~(\ref{pot}). This interaction is  crucial for the axion mechanism,
 because it transmits the breaking of the global symmetry triggered by the vev of $S$ to the Higgs sector.
One or both of the Higgs doublets can remain light, at the electroweak scale.
The presence of an instability is subject to the details of the two-Higgs potential~\cite{THDM}, but this does not change the essential point. Independently of the model implementation, the field $S$ containing the invisible axion 
$a=\sqrt{2} {\rm Im }\, S$ with large decay constant $f_a \approx \langle S\rangle$
is a perfect candidate to play the role of the field $S$ in our Higgs stabilization mechanism. 

\smallskip

KSVZ axion models~\cite{KSVZ} use a single Higgs doublet and a complex scalar $S$ coupled to new heavy vector-like fermions $\Psi$.
The Dirac mass term $M\,\bar\Psi \Psi$ is forbidden by imposing the symmetry
\be \Psi_L\to -\Psi_L, \qquad \Psi_R\to \Psi_R,\qquad S \to - S.\ee
Then the mass
of the heavy fermions comes  only from the vev of $S$:
\be 
\lambda_\Psi~S \bar\Psi \Psi  + V(H,S) .\ee
The resulting model has a spontaneously broken U(1) global symmetry
\be \Psi \to e^{i\gamma_5 \alpha}\Psi,\qquad S\to e^{-2i\alpha} S\ee
which gives rise to a light axion $a=\sqrt{2} {\rm Im }\, S$ with large decay constant $f_a \approx \langle S\rangle$.
The scalar potential of the theory is precisely of the form in eq.~(\ref{pot}),
although the coupling $\lambda_{HS}$ plays no role in axion phenomenology
because both $|S|^2$ and $|H|^2$ are separately invariant under the global symmetry.

The decay constant of the axion is allowed to lie in the range
\be 10^9\,{\rm GeV} < f_a  < 10^{12}\,{\rm GeV}\ . \label{farange}\ee
The lower bound comes from non-observation of axion emission from stars and supernov\ae{}.
The upper bound comes from requiring that the axion dark matter density 
\be \Omega_a \approx 0.15  \left(\frac{f_a}{10^{12}\,\rm GeV}\right)^{7/6}\left(\frac{a_*}{f_a}\right)^2\ee
does
not exceed the observed value $\Omega_{\rm DM} \approx 0.23$
under the assumption that the axion vev $a_*$
in the early universe
was of the order of $f_a$~\cite{axionDM}.
The resulting range of singlet mass $M_S$, which we can roughly take to be the same as the range for $f_a$ in eq.~(\ref{farange}), overlaps with the range that can stabilize the SM Higgs potential from $m_h\simgt 119$ GeV  (for $M_S\sim 10^9$ GeV) to $m_h\simgt 124$ GeV (for $M_S\sim 10^{12}$ GeV),
as can be inferred from fig.~\ref{SMLIMH}.

\subsection{Unitarized Higgs Inflation}

In the original proposal for SM Higgs inflation \cite{higgsinf}, a large non-minimal coupling $\xi\sim 10^4$ of the Higgs boson to gravity, $\xi H^\dagger H R$, drives inflation at large Higgs field values. However, due to the same large non-minimal coupling, the unitarity cutoff \cite{unitarity} becomes $\Lambda_{\rm SM}=M_{\rm Pl}/\xi$ which is not only much lower than the Planck scale but also lower than the field scales at which the inflationary plateau develops. Although the unitarity cutoff during inflation could be larger than the one in the vacuum \cite{bkgdep}, the perturbative expansion is still questionable and the very existence of an inflationary plateau beyond $\Lambda_{\rm SM}$ is jeopardized. In addition to this unitarity problem, 
in all the Higgs mass range allowed by ATLAS and CMS, the potential instability
that develops at large field values could also destroy the flat plateau induced by gravitational effects \cite{EEGIRS}. 

To fix the unitarity problem, extra dynamical degrees of freedom are required to restore unitarity without ruining the flat plateau. A UV complete model with a real singlet scalar of sigma-model type was proposed in~\cite{gianlee} and it is natural to ask if this singlet could also solve simultaneously the instability problem.
The Jordan-frame Lagrangian of the model is
\bea
\frac{{\cal L}_J}{\sqrt{-g_J}}&=&\frac{1}{2} \Big(M^2+\xi\sigma^2+2\zeta  H^\dagger H\Big)R-\frac{1}{2}(\partial_\mu\sigma)^2-|D_\mu H|^2 \nonumber \\
&&-\frac{1}{4}\lambda_\sigma \Big(\sigma^2- \Lambda^2+2\frac{\lambda_{H\sigma}}{\lambda_\sigma} H^\dagger H\Big)^2-\Big(\lambda_H-\frac{\lambda^2_{H\sigma}}{\lambda_\sigma}\Big) \Big(H^\dagger H-\frac{v^2}{2}\Big)^2 \ ,
\label{jordanaction0}
\eea
where $M,\Lambda$ and $v$ are mass parameters with $v\ll M,\Lambda$ (so that the $\sigma$ field is heavy)
and $\xi,\zeta$ are positive non-minimal couplings with $\xi\gg \zeta$. 
The scalar potential derived from eq.~(\ref{jordanaction0}) falls in the class of models considered in this paper, once we identify $\sigma$ with $S$ (although the singlet now is real). In what follows we will also use the effective quartic coupling defined as
\be
\lambda\equiv \lambda_H-\frac{\lambda^2_{H\sigma}}{\lambda_\sigma}\ .
\ee

The large nonzero vev of $\sigma$, $\langle\sigma\rangle\simeq \Lambda$, is crucial to make the unitarity cutoff $\Lambda_{UV}$ larger. It is straightforward to find that 
\be
\Lambda_{UV}=  \Big(1+6r \xi\Big)\frac{M_{\rm Pl}}{\xi}\ ,
\label{UVcutoff}
\ee
where the Planck mass is now $M_{\rm Pl}^2= M^2+\xi\Lambda^2$, and we measure the contribution of the $\sigma$ vev by the ratio $r= \xi \Lambda^2/M_{\rm Pl}^2$, which in general can take values from 0 to 1. One can see how for a negligible vev ($r\rightarrow 0$, as was the case of the SM Higgs inflation case) the cutoff is $M_{\rm Pl}/\xi$ while it is pushed up to $r M_{\rm Pl}$ for moderate values of $r\simgt 1/\xi$.

\smallskip

As described in \cite{gianlee}, in this scenario the sigma field dominates inflation due to its large non-minimal coupling to $R$ while the Higgs field follows the sigma field along a flat direction, provided $\lambda_{H\sigma}<0$. In fact, in unitary gauge with $H^T=(0,h)/\sqrt{2}$, for $\sigma, h\gg M/\sqrt{\xi},\Lambda$, slow-roll inflation takes place along $h\approx \sqrt{(-\lambda_{H\sigma}/\lambda_H)}\, \sigma$, while the direction orthogonal to the inflaton gets mass of order $\sqrt{-2\lambda_{H\sigma}}\,M_{\rm Pl}/\sqrt{\xi}$ (identified with a larger background-dependent cutoff in the Higgs inflationary regime \cite{gianlee}). Here, a negative $\lambda_{H\sigma}$ is crucial for the sigma model to reproduce the Higgs inflation model below the sigma mass and for the Higgs boson to participate in inflaton dynamics beyond $\Lambda$.
Then, the Einstein-frame potential along the inflaton field direction $\chi=\sqrt{6} M_{\rm Pl}\ln(\sqrt{\xi}\sigma/M_{\rm Pl})$, is given by 
\be
V(\chi)\simeq \frac{\lambda_\sigma \lambda}{4\lambda_H}\,\frac{M^4_{\rm Pl}}{\xi^2} \Big(1-2\,e^{-2\chi/\sqrt{6} \,M_{\rm Pl}}\Big)\ .
\label{Vinfl}
\ee
In order to match the COBE normalization of the power spectrum, we obtain the following condition on the parameters,
\be
\frac{\sqrt{\lambda_\sigma}}{\xi}=2\times 10^{-5}\sqrt{\frac{\lambda_H}{\lambda}}\ ,
\label{COBE}
\ee
which explains the need of having $\xi\gg 1$.

The mass of the $\sigma$ field in the vacuum, on the other hand, is given by
\be
M_{\bar\sigma}^2=\lambda_\sigma \frac{2 r M_{\rm Pl}^2}{(1+6r\xi)\xi}\simeq\lambda_\sigma\frac{M_{\rm Pl}^2}{3\xi^2}\ ,
\ee
(where $\bar\sigma$ denotes the canonically normalized field, and we have required $r\simgt 1/\xi$ for the last expression), which is $r$-independent and of the same order of $\Lambda_{\rm SM}$, the UV cutoff for $r=0$.
The $\sigma$ field could be lighter than $\Lambda_{\rm SM}$ at the cost of reducing $r$ below $1/\xi$, but then the unitarity cutoff (\ref{UVcutoff}) would decrease, back to its SM value. In principle, one could also lower the sigma mass below the unitarity cutoff by choosing a small value of $\lambda_\sigma$. However, the COBE constraint (\ref{COBE}) precisely fixes the sigma mass in the vacuum to be (taking $\lambda_H/\lambda$ of order 1)
\be
M_{\bar\sigma}\approx 10^{13}\,{\rm GeV} \ .
\ee
As we would like the singlet to be lighter than $\Lambda_I$ in order to cure that instability problem, we are forced to a region of Higgs masses for which $\Lambda_I> 10^{13}$ GeV. From fig.~\ref{SMLIMH} we see that this requires $m_h>125$ GeV (at 90\% CL in $M_t$ and $\alpha_{\rm s}$), which is marginally compatible with
the ATLAS and CMS hint of $m_h\sim 124$--126 GeV.

\medskip

As   explained above, the scenario requires $\lambda_{H\sigma}<0$, so that the stability condition
to be satisfied (at all scales) is $\lambda=\lambda_H-\lambda^2_{H\sigma}/\lambda_\sigma>0$. As can be seen from eq.~(\ref{Vinfl}) this condition amounts to requiring a positive vacuum energy during inflation.
From our general analysis we know that in this case the tree-level threshold effect alone would not improve the vacuum stability and RG effects above the sigma mass are necessary for stabilizing the potential. The instability scale $\Lambda_I$ should be pushed up, if not all the way to the Planck scale, at least sufficiently high to maintain the inflationary plateau stable in an interval that provides enough e-folds of expansion. Such interval
is approximately $\Lambda<\sigma < 10\Lambda$, so that it would be sufficient to increase $\Lambda_I$
by one or two orders of magnitude.

Concerning the RG evolution of couplings above the sigma-field threshold, we follow \cite{wilczek}
and take into account the effects of a non-minimal coupling to gravity of the sigma and Higgs fields
through suppression factors in the RGEs, which become:
\bea
(4\pi)^2\frac{d\lambda _{{H}}}{d\ln\mu} &=& (12 y_t^2-3 {g'}^2-9 g^2) \lambda _{{H}}-6 y_t^4+\frac{3}{8}\left[2g^4+ ({g'}^2+g^2)^2\right]  \nonumber \\
&&+(18c^2_h+6) \lambda _{{H}}^2+2c^2_\sigma \lambda _{{H\sigma}}^2,  \nonumber \\
(4\pi)^2\frac{d\lambda _{{H\sigma}}}{d\ln\mu} &=& \frac{1}{2}\lambda _{{H\sigma}} \left[12 y_t^2-3g^{\prime 2}-9g^2+12(1+c^2_h) \lambda _{{H}}+12c^2_\sigma \lambda _{{\sigma}}\right]+8 c_h c_\sigma\lambda _{{H\sigma}}^2, \\
(4\pi)^2\frac{d\lambda _{{\sigma}}}{d\ln\mu} &=& 2(3+c^2_h) \lambda _{{H\sigma}}^2+18 c^2_\sigma\lambda _{{\sigma}}^2 \ . \nonumber
\eea
(Numerical factors are now appropriate for a real scalar $\sigma$, and therefore these
equations do not reduce to~(\ref{RG}) in the limit $c_{\sigma,h}\to 1$).
The suppression factors, $c_\sigma$ and $c_h$, are given in terms of the Weyl rescaling factor 
$\Omega^2=(M^2/M^2_{\rm Pl})(1+\xi\sigma^2/M^2)$ as
$c_\sigma= \Omega^{-2} (\partial\chi/\partial \sigma)^{-2}\simeq (1+M^2/\xi\sigma^2)/(6\xi)\ll 1$ and
$c_h= \Omega^{-2} (\partial\phi/\partial h)^{-2}=1$, where $\chi$ and $\phi$ are canonical fields.
Therefore, the running of $\lambda_H$ is SM-like, as loops containing the sigma scalar are suppressed for $c_\sigma \ll 1$ and Higgs loops are the same as in the SM. Then, we use the above RG equations to get the corresponding equation for $\lambda\equiv \lambda_H-\lambda^2_{H\sigma}/\lambda_\sigma$, with $\delta\lambda \equiv \lambda^2_{H\sigma}/\lambda_\sigma$, as
\bea
\frac{d\lambda }{d\ln\mu} \approx \beta^{\rm SM}_\lambda
+\frac{8}{(4\pi)^2}(3\lambda+\delta\lambda)\,\delta\lambda\,.
\eea
Consequently, as compared to the SM, the sigma scalar gives an extra positive contribution in $\beta_\lambda$. Therefore, the loop effects of the sigma scalar couplings can make the instability scale larger than the SM one for a sizable value of $\delta\lambda$ at the sigma mass scale.
In conclusion, unitarized Higgs inflation could still be a viable possibility without an instability problem.

\section{Conclusions}

The recent LHC Higgs searches have narrowed down the Higgs mass to a range which most likely leads to a large-field instability of the potential, in the context of the SM. One can view this result as an opportunity to learn new information about the early universe through collider experiments~\cite{gesp}, or as the need for a UV modification of the SM. 

The lack of early discoveries at the LHC is a motivation for considering minimal, albeit technically unnatural, theoretical descriptions of the particle world. In this paper we have analyzed the most minimal UV modification that eliminates the EW vacuum instability, adding one singlet scalar field to the SM degrees of freedom. As the theory is renormalized to low energies and the new singlet is integrated out, the Higgs quartic coupling is shifted downwards, making the Higgs mass smaller. As a result, the physical Higgs mass lies inside the region of instability in the pure SM, although there is no actual instability in the complete UV theory. From this point of view, the instability for $m_h$ in the allowed range 115--127~GeV is just a mirage.

The proposed stabilization mechanism relies on a threshold correction to the Higgs quartic coupling, whose size is independent of the singlet mass. The necessary ingredients are a singlet self-quartic coupling ($\lambda_S$), a mixed quartic coupling with the Higgs ($\lambda_{HS}$) and a non-zero vev for the singlet. The mechanism can be operative even for a very heavy singlet, as long as its mass is smaller than the instability scale $\Lambda_I$. Occurring at tree level, the effect is sizable and robust.

The analysis of the effect involves some subtleties because, at the singlet threshold, both the Higgs quartic coupling and the stability conditions are shifted by the same amount. We have shown that, when $\lambda_{HS}$ is positive, the stability conditions become weaker as the field value is increased above the singlet mass. In this situation, the tree-level contribution is very effective in stabilizing the potential. On the other hand, for negative $\lambda_{HS}$, the shifts in the Higgs quartic coupling and in the stability condition essentially cancel out, and one has to rely on RG effects. These can help the stabilization, but larger singlet couplings are needed to obtain the desired effect.

The minimal modification of the SM that we have considered, with the addition of one singlet scalar, has motivations that are independent of the stability of the EW vacuum. We have investigated three examples. The new singlet can set the scale of the right-handed neutrino mass in the sea-saw mechanism; or it could play the role of the invisible axion; or finally it could unitarize models with large gravitational non-minimal couplings of the Higgs field invoked for inflationary dynamics. In each case, we were able to define the range of Higgs masses for which the corresponding singlet could also be used to stabilize the SM Higgs potential. We find that stabilization is possible for 
any Higgs mass allowed by the current LHC limits, in the case of leptogenesis with dominant initial right-handed neutrino density; for
$m_h> 119$~GeV, in the case of leptogenesis with thermal initial right-handed neutrino density or in the case of the invisible axion; for $m_h> 120$~GeV, in the case of leptogenesis with vanishing initial right-handed neutrino density; for $m_h> 125$~GeV in the case of unitarized Higgs inflation.

\section*{Acknowledgments}
J.R.E. thanks J.L.F. Barb\'on for interesting discussions.
J.E.-M. and J.R.E. acknowledge CERN for hospitality and partial financial support. 
This work was supported by the ESF grant MTT8; by SF0690030s09 project; the Spanish
Ministry MICINN under contracts FPA2010-17747 and FPA2008-01430;
the Spanish Consolider-Ingenio 2010 Programme CPAN (CSD2007-00042);
and the Generalitat de Catalunya grant 2009SGR894.




\end{document}